\documentstyle{article}
\newread\epsffilein    
\newif\ifepsffileok    
\newif\ifepsfbbfound   
\newif\ifepsfverbose   
\newdimen\epsfxsize    
\newdimen\epsfysize    
\newdimen\epsftsize    
\newdimen\epsfrsize    
\newdimen\epsftmp      
\newdimen\pspoints     
\def\epsfbox#1{\global\def\epsfllx{72}\global\def\epsflly{72}%
   \global\def\epsfurx{540}\global\def\epsfury{720}%
   \def\lbracket{[}\def\testit{#1}\ifx\testit\lbracket
   \let\next=\epsfgetlitbb\else\let\next=\epsfnormal\fi\next{#1}}%
\def\epsfgetlitbb#1#2 #3 #4 #5]#6{\epsfgrab #2 #3 #4 #5 .\\%
   \epsfsetgraph{#6}}%
\def\epsfnormal#1{\epsfgetbb{#1}\epsfsetgraph{#1}}%
\def\epsfgetbb#1{%
\openin\epsffilein=#1
\ifeof\epsffilein\errmessage{I couldn't open #1, will ignore it}\else
   {\epsffileoktrue \chardef\other=12
    \def\do##1{\catcode`##1=\other}\dospecials \catcode`\ =10
    \loop
       \read\epsffilein to \epsffileline
       \ifeof\epsffilein\epsffileokfalse\else
          \expandafter\epsfaux\epsffileline:. \\%
       \fi
   \ifepsffileok\repeat
   \ifepsfbbfound\else
    \ifepsfverbose\message{No bounding box comment in #1; using defaults}\fi\fi
   }\closein\epsffilein\fi}%
\def\epsfclipstring{}
\def\epsfsetgraph#1{%
   \epsfrsize=\epsfury\pspoints
   \advance\epsfrsize by-\epsflly\pspoints
   \epsftsize=\epsfurx\pspoints
   \advance\epsftsize by-\epsfllx\pspoints
   \epsfxsize\epsfsize\epsftsize\epsfrsize
   \ifnum\epsfxsize=0 \ifnum\epsfysize=0
      \epsfxsize=\epsftsize \epsfysize=\epsfrsize
      \epsfrsize=0pt
     \else\epsftmp=\epsftsize \divide\epsftmp\epsfrsize
       \epsfxsize=\epsfysize \multiply\epsfxsize\epsftmp
       \multiply\epsftmp\epsfrsize \advance\epsftsize-\epsftmp
       \epsftmp=\epsfysize
       \loop \advance\epsftsize\epsftsize \divide\epsftmp 2
       \ifnum\epsftmp>0
          \ifnum\epsftsize<\epsfrsize\else
             \advance\epsftsize-\epsfrsize \advance\epsfxsize\epsftmp \fi
       \repeat
       \epsfrsize=0pt
     \fi
   \else \ifnum\epsfysize=0
     \epsftmp=\epsfrsize \divide\epsftmp\epsftsize
     \epsfysize=\epsfxsize \multiply\epsfysize\epsftmp   
     \multiply\epsftmp\epsftsize \advance\epsfrsize-\epsftmp
     \epsftmp=\epsfxsize
     \loop \advance\epsfrsize\epsfrsize \divide\epsftmp 2
     \ifnum\epsftmp>0
        \ifnum\epsfrsize<\epsftsize\else
           \advance\epsfrsize-\epsftsize \advance\epsfysize\epsftmp \fi
     \repeat
     \epsfrsize=0pt
    \else
     \epsfrsize=\epsfysize
    \fi
   \fi
   \ifepsfverbose\message{#1: width=\the\epsfxsize, height=\the\epsfysize}\fi
   \epsftmp=10\epsfxsize \divide\epsftmp\pspoints
   \vbox to\epsfysize{\vfil\hbox to\epsfxsize{%
      \ifnum\epsfrsize=0\relax
        \includegraphics{#1}%
      \else
        \epsfrsize=10\epsfysize \divide\epsfrsize\pspoints
        \includegraphics{#1}%
      \fi
      \hfil}}%
\global\epsfxsize=0pt\global\epsfysize=0pt}%
{\catcode`\%=12 \global\let\epsfpercent=
\long\def\epsfaux#1#2:#3\\{\ifx#1\epsfpercent
   \def\testit{#2}\ifx\testit\epsfbblit
      \epsfgrab #3 . . . \\%
      \epsffileokfalse
      \global\epsfbbfoundtrue
   \fi\else\ifx#1\par\else\epsffileokfalse\fi\fi}%
\def\epsfempty{}%
\def\epsfgrab #1 #2 #3 #4 #5\\{%
\global\def\epsfllx{#1}\ifx\epsfllx\epsfempty
      \epsfgrab #2 #3 #4 #5 .\\\else
   \global\def\epsflly{#2}%
   \global\def\epsfurx{#3}\global\def\epsfury{#4}\fi}%
\def\epsfsize#1#2{\epsfxsize}
\def\pixdir{.}

\def\twofactor{.4}
\def\threefactor{.27}

\def\pixtype#1{\let\epsfinsert=\epsfbox[#1]}

\def\captionfont{\small}

\def\picture#1#2#3{\begin{tabular}{c}%
\mbox{\epsfxsize=#1\epsfinsert{\pixdir/#2}}\\\mbox{\captionfont #3}\end{tabular}}

\def\pich#1#2{\raisebox{-0.5\baselineskip}{\epsfysize=#1\epsfinsert{\pixdir/#2}}}

\def\pic#1#2{\mbox{\epsfxsize=#1\hsize%
\epsfinsert{\pixdir/#2}}}

\def\twoinrow#1#2#3#4{%
\mbox{\picture{\twofactor\hsize}{#1}{#2}\hspace{.05\hsize}%
\picture{\twofactor\hsize}{#3}{#4}}}

\def\threeinrow#1#2#3#4#5#6{%
\mbox{\picture{\threefactor\hsize}{#1}{#2}\hspace{.04\hsize}%
\picture{\threefactor\hsize}{#3}{#4}\hspace{.04\hsize}%
\picture{\threefactor\hsize}{#5}{#6}}}

\def\fplotsq#1#2{\mbox{\epsfxsize=#1\hsize%
\epsfbox[65 10 580 535]{\pixdir/#2}}}

\let\epsfinsert=\epsfbox

\begin{document}
\begin{titlepage}
\vbox{\hfill UM-P-93/05}
\begin{center}
\vspace{4cm}
{\Large\bf Fermionic Field Theory and \\}
{\Large\bf Gauge Interactions on Random Lattices\\}
\vspace{1cm}
{C. J. Griffin and T. D. Kieu,}\\
\vspace{1cm}
School of Physics,\\
University of Melbourne,\\
Parkville VIC 3052,\\
AUSTRALIA\\
\vspace{2cm}
\begin{quotation}
Random-lattice fermions have been shown to be free of the doubling problem
if there are no interactions or interactions of a non-gauge nature.
However, gauge interactions impose stringent constraints
as expressed by the Ward-Takahashi identities which could
revive the
free-field suppressed doubler modes in loop diagrams.  After introducing
a formulation for fermions on a new kind of random lattice, we compare
random, naive and Wilson fermions in two dimensional Abelian background gauge
theory. We show that the doublers are revived for random lattices in the
continuum limit, while demonstrating that gauge invariance plays the critical
role in this revival.
Some implications of the persistent doubling phenomenon
on random lattices are also discussed.
\end{quotation}
\end{center}
\end{titlepage}
\topmargin=0in


\section{Introduction}
Lattice regularisation is particularly important not only because
of its fundamental and traditional role in {\it defining} and
regularising quantum field theory but also because it opens
the way to rigorous, non-perturbative treatments.  However,
on the lattice chiral fermions suffer the doubling phenomenon
of generation of extraneous species \cite{NN} in such a
way that the net chirality is zero.  This long-standing doubling problem
is the most important problem of quantum field theory.
More than just technical, it is inherently fundamental
as emphasised by several no-go theorems, and as the
difficulty in defining Chiral Gauge Theory is shared by {\it all} other
regulators in one way or another.

The doubling problem of lattice fermions is inevitable according to the
Nielsen Ninomiya no-go theorem \cite{NN} if the bilinear free-field action satisfies
the conditions of reflection positivity, locality, global axial symmetry,
and translational invariance at a fixed scale.
An obvious resolution of the doubling problem is thus to relax one of those
conditions to obtain, in the order listed above,
non-hermitian \cite{non-herm}, non-local \cite{non-local},
Wilson \cite{wilson}, or random-lattice \cite{ran,ran2,espiru,espiru2,chiu}
fermion formulations.
These formulations are all free of doublers when there are no
interactions or when the interactions are of a non-gauge
nature \cite{peran,tdkthesis}: the extra poles in the propagators
are removed as the lattice spacing $a$ decreases, leaving a single fermion mode
in the continuum limit.

Gauge interactions behave very differently on account of a
unique and special property.  Local gauge invariance imposes severe
constraints on the theory, expressed mathematically in the Ward-Takahashi
identities.  In particular, for electrodynamics, the fermion-gauge vertex $\Lambda_\mu$ is related to the
free inverse propagator $G^{-1}_0$ at zero momentum transfer,
\begin{eqnarray}
\Lambda_\mu(p,p)=-g\frac{\partial}{\partial p^\mu}G^{-1}_0(p),
\end{eqnarray}
giving the interaction vertices mode dependency. Modifications made to the action
$\overline\Psi G\Psi$ must be compensated by an appropriate well prescribed change
of conserved current, and thus vertex, in order to respect gauge invariance and the minimal coupling prescription.
Such modifications have been shown to revive the doubled modes
in studies of some non-local
\cite{bodwin} and non-hermitian
formulations \cite{kashiwa,tdkprep}, even though
those modes are suppressed at the  free-field level.
For this reason, we investigate the issue of fermion
doubling on random lattices with gauge interactions \cite{wheater}.

In random lattice approaches, suitable quantities are measured
on a random lattice then averaged, either quenchedly or annealedly,
over an ensemble of lattices.  Apart from the extra work involved in
generating an ensemble of random lattices, this approach better approximates
the scale-free rotational and translational symmetry of the
continuum than regular lattices.  Thus, the continuum limit
may be more easily reached on random lattices than on regular
lattices of the same size.  More relevant to this discussion,
since there is no
fixed Brillouin zone, there need be no extra poles of the
propagator. Even if extra poles do exist, the one-to-one correspondence between
propagator poles in momentum space and zero modes is not necessarily valid
since plane waves are
no longer eigenstates of the Dirac operator. Alternatively, one could appeal to
the fact that there is no transfer matrix on a random lattice (at least
for a finite lattice) since there are no identical timeslices, to
argue that there may not be a clear relation between poles of the
inverse propagator and the particle spectrum \cite{espiru2}.

This expectation of no doubling on random lattices has been realised
in various studies of free-field theory in both two and
four dimensions \cite{ran2,espiru}.  It has similarly been shown that random lattice
theories with four-point interactions are also doubler free \cite{peran}.
The full gauge invariant formulation has not yet been properly considered on account of
problems associated with identifying
the appropriate conserved gauge current that appears in the action.
Even though we will not consider 
this current explicitly, we consider a
quantity which automatically includes the correct current. 
We focus on the connected gauge field two-point function, which can be computed directly
from the action, 
\begin{eqnarray}
-\ln{\rm Det\,}(G_AG_0^{-1})&=&\Gamma_2+O(g^4),\nonumber\\
\Gamma_2&=&{\rm Tr}[(G_A^{-1}G_0-1)-\frac{1}{2}(G_A^{-1}G_0-1)^2]\\
&=&\pich{18pt}{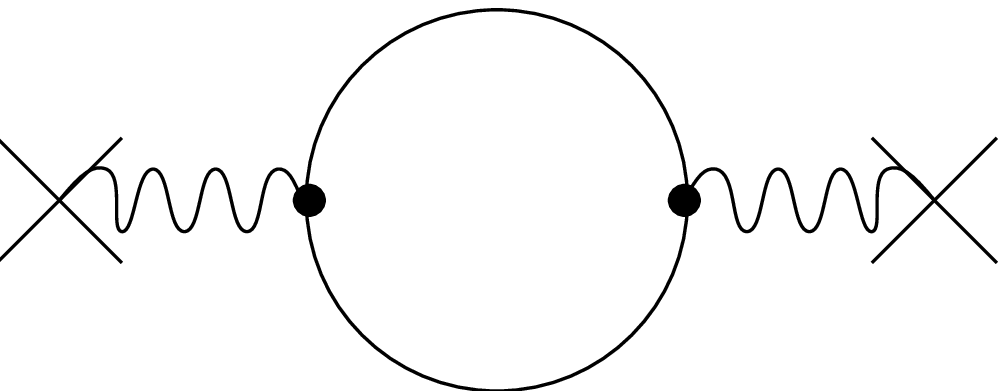}+\pich{18pt}{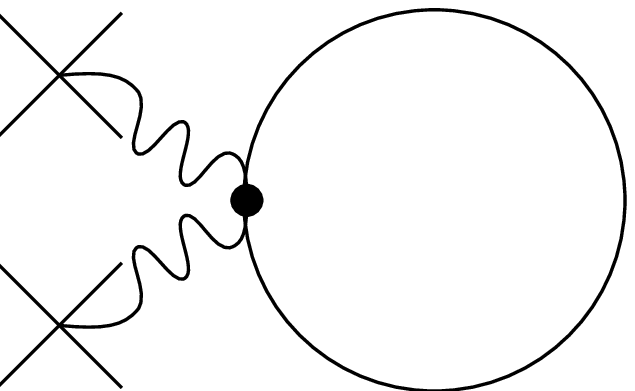}+\ldots,\nonumber
\end{eqnarray}
where $G_A^{-1}$ is the fermion propagator in a gauge field
$A_\mu$ and $g$ is the gauge coupling.
On the lattice, $\Gamma_2$ is the required two-point function up to a numeric
constant and 
$O(A^4g^4a^2)$ correction terms. Choosing a small coupling $gA_\mu\ll1$
keeps this correction under control even on a finite lattice.
We choose to study two dimensional background QED, in which
there are no internal photon lines in the loops.
Hence, we expect to see a clean signal which simply counts the number of fermion modes
in the continuum $(a=0)$ limit.
A comparison with identical calculations for naive and Wilson fermions on 
two dimensional square lattices,
which are known to be four-fold doubling and doubler-free respectively,
over a range of $a$, clarifies the continuum limit behaviour of our random lattices. After introducing the random lattice and fermion action used in this
work and comparing with other random lattices, we verify that the free-field case does indeed have doubler suppression.
However, in the fully interacting gauge invariant theory, the two-point function does not 
suppress doubling. If gauge invariance is broken on the lattice, the 
doubled modes are eliminated.

\begin{figure}
\begin{center}
\pic{0.4}{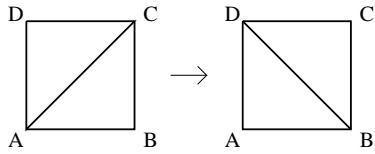}
\end{center}
\caption{\label{fig.alex}An Alexander `flip'}
\end{figure}

\section{Random lattice construction}
Random lattices are constructed using a method borrowed from studies of
simplicial gravity: an initial regular triangulated array of $N$ fixed square lattice
vertices is rearranged by a sequence of Alexander `flip' moves \cite{alexander,fractaldim}. Figure~\ref{fig.alex}
pictorially demonstrates a single flip:
A quadrilateral $ABCD$, with a unique internal link $AC$ is randomly chosen,
the internal link is deleted, and a new link $BD$ is introduced.
In simplicial gravity studies there is no restriction on the local curvature, 
and bond lengths are generally 
kept fixed. This flip prescription
is sufficient to build an arrangement of links and vertices which is 
statistically uneffected by further flipping \cite{fractaldim}. In the case at hand, we require
the local curvature to be zero. The most straight-forward way to accomidate
this is to fix the vertex positions and allow the links to have varying lengths.
A further condition must be included: a flip is performed only if the planar local orientability of the links 
\begin{figure}
\begin{center}
\threeinrow{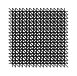}{initial}{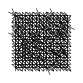}{$1$ scan}{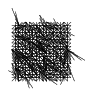}{$8$ scans}
\end{center}\caption{\label{fig.lat}A typical $256$ vertex lattice}
\end{figure}
is preserved. This prevents undesirable two dimensional behaviour such as crossed links (which correspond to parts of the lattice overlapping each other), and zero area simplices from being generated in the flipping procedure.
Both the number of vertices and the number of links can be
fixed on the lattice. 
This fixed-vertex construction has several features: 
the entire process is $O(N)$, each random lattice 
has a fixed size independent of the exact details of the 
randomising procedure, so measured quantities
do not need to be scaled by the average link length $s$, and the scale
below which the lattice ceases to be a meaningful 
representation of space-time is cleanly identified by $s=a$.
We consider a quantity of flips called a `scan', which is defined as $6N$
successful flips. This measure of flipping is independent of lattice size,
and allows the generic behaviour of the flipping procedure to be uncovered
in more detail.
At first glance, it may seem reasonable that the flipping process leads to
equilibrated
configurations, 
however it is not clear that such an equilibrated structure
actually exists, or is useful in these studies.

Beginning from the regular configuration, the links are in their shortest possible state. Random flipping will thus tend to increase the link length.
In fact, this behaviour persists for a realtively small number of 
scans even after the initial structure has been
well destroyed.
\begin{figure}
\begin{center}
\twoinrow{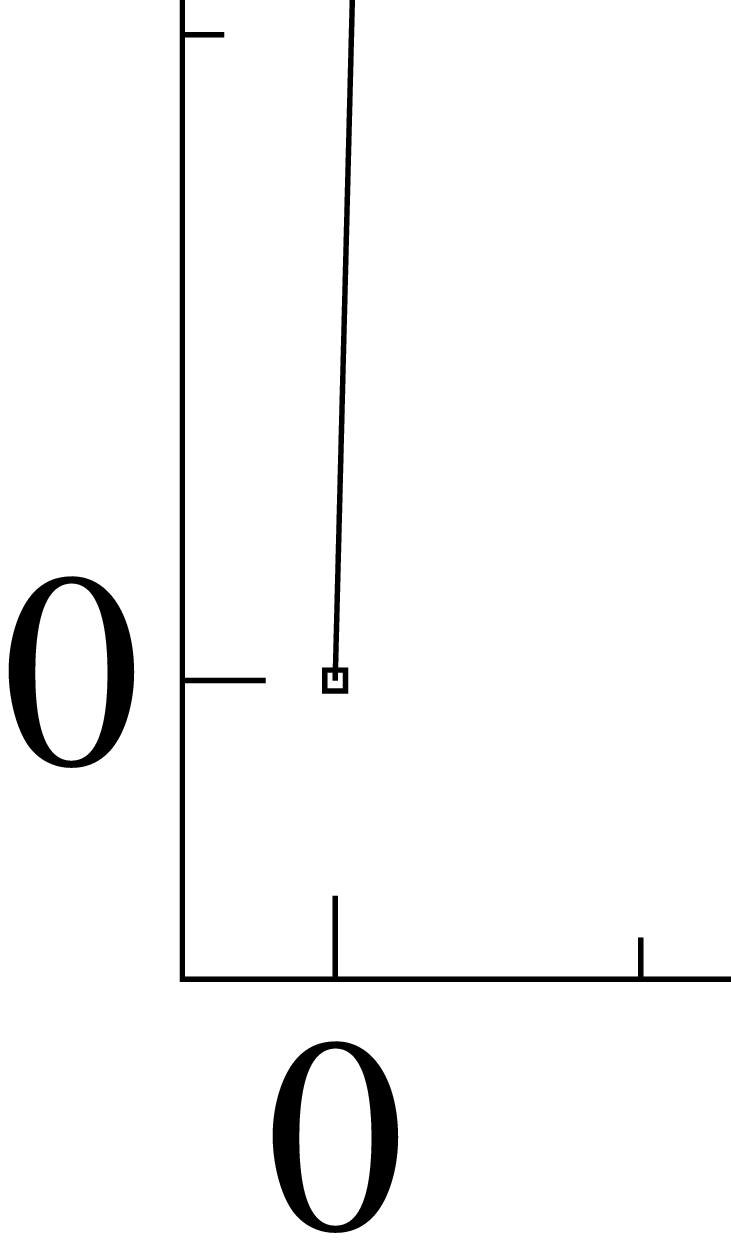}{\ref{fig.linklength}a. $s$ increases with $n$, universally.}%
{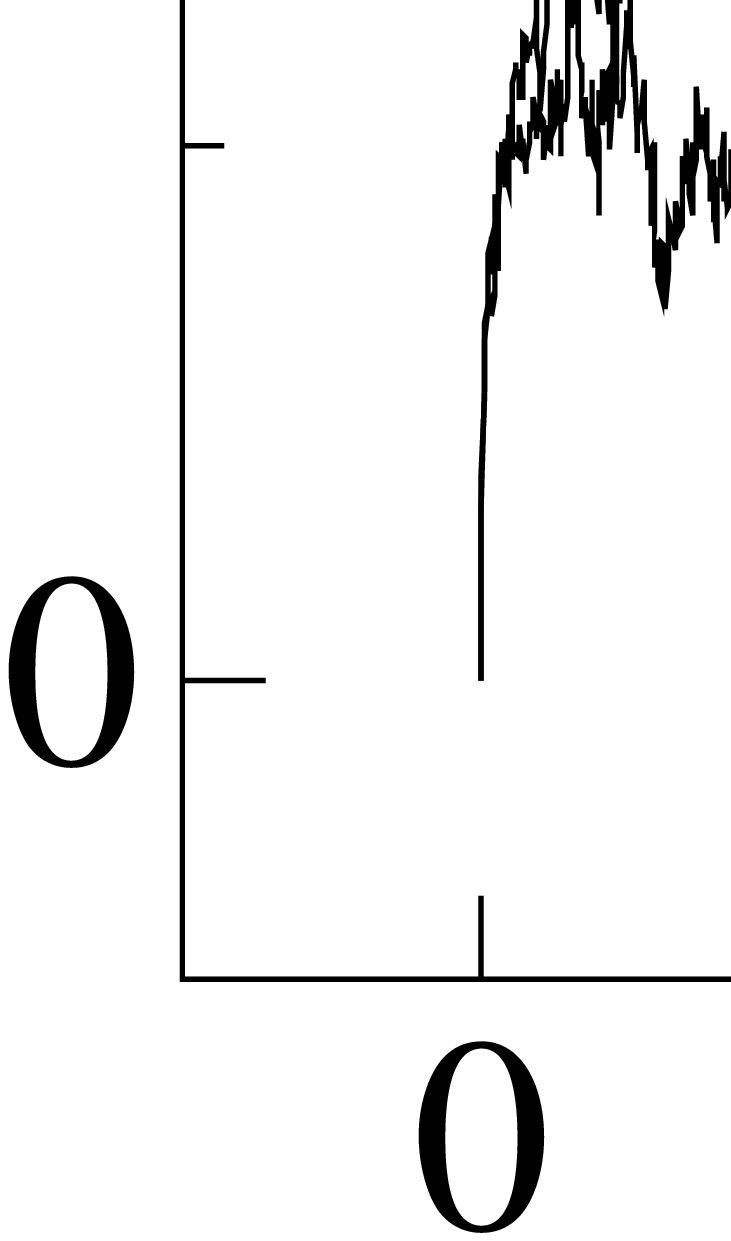}{\ref{fig.linklength}b. Flip stability.}
\end{center}
\caption{\label{fig.linklength}Lattice flipping.}
\end{figure}
Figure~\ref{fig.linklength}a. demonstrates the behaviour
of $\Delta s=s(n)-s(0)$, the average link extension as a function of 
scan number $n$, for various lattice sizes, clearly demonstrating the 
universal nature of the randomising process. 
The link extension is numerically approximated by
\begin{equation}
(\Delta s)^2=\frac{1}{2}\;\log\;n,
\end{equation}
with fluctuations about this value diminishing as the lattice size is increased.

One may expect there would be some retarding of the growth;
as the length of a link increases, the paths of other growing links
are blocked, which may lead to a grid--lock situation, where the structure is
equilibrated to a certain extent. 
That this equilibration is a global phenomenon is easily understood by 
contemplating the behaviour on larger lattices, where longer links are 
required in order to introduce the same effect. Thus the equilibrium, if it
exists,
is probably induced by a finite size effect. Such notions have already been 
observed in the quantum gravity case
where the fractal dimension of the equilibrated structure in two dimensions
diverges with $\log N$ \cite{fractaldim}, indicating that the establishment of 
equilibrium is intimately connected with the global lattice size.

Since the vertices of this lattice are anchored and we have chosen toroidal 
boundary conditions, another scenario is possible; a fault may form which 
wraps abound the entire torus, link growth perpendicular to the fault is prevented, and links within the fault cannot be flipped out, thus the fault is 
more difficult to remove when it gets longer, and likely to grow further. 
This leads to a runaway situation, which is once again related to the 
finite lattice size.

Figure~\ref{fig.linklength}b. demonstrates both the equilibrated, and runaway situation;
initial geometries are identical, only the random seed has been changed.
In order to avoid these undesirable global effects, we choose to flip
as little as possible,
randomising with $1\ldots8$ scans, see Fig.~\ref{fig.lat} for a typical lattice.
By construction, the lattice violates translational invariance
and is thus immediately suitable for our purpose. 

It is interesting to compare
with other random lattice formulations, specifically the CFL lattice of 
ref.\ \cite{ran}.
For these purposes, we consider the coordination number distribution, $\rho_N(C)$
and the link length distribution $\rho_L(x)$.

\begin{figure}
\begin{center}
\pic{0.7}{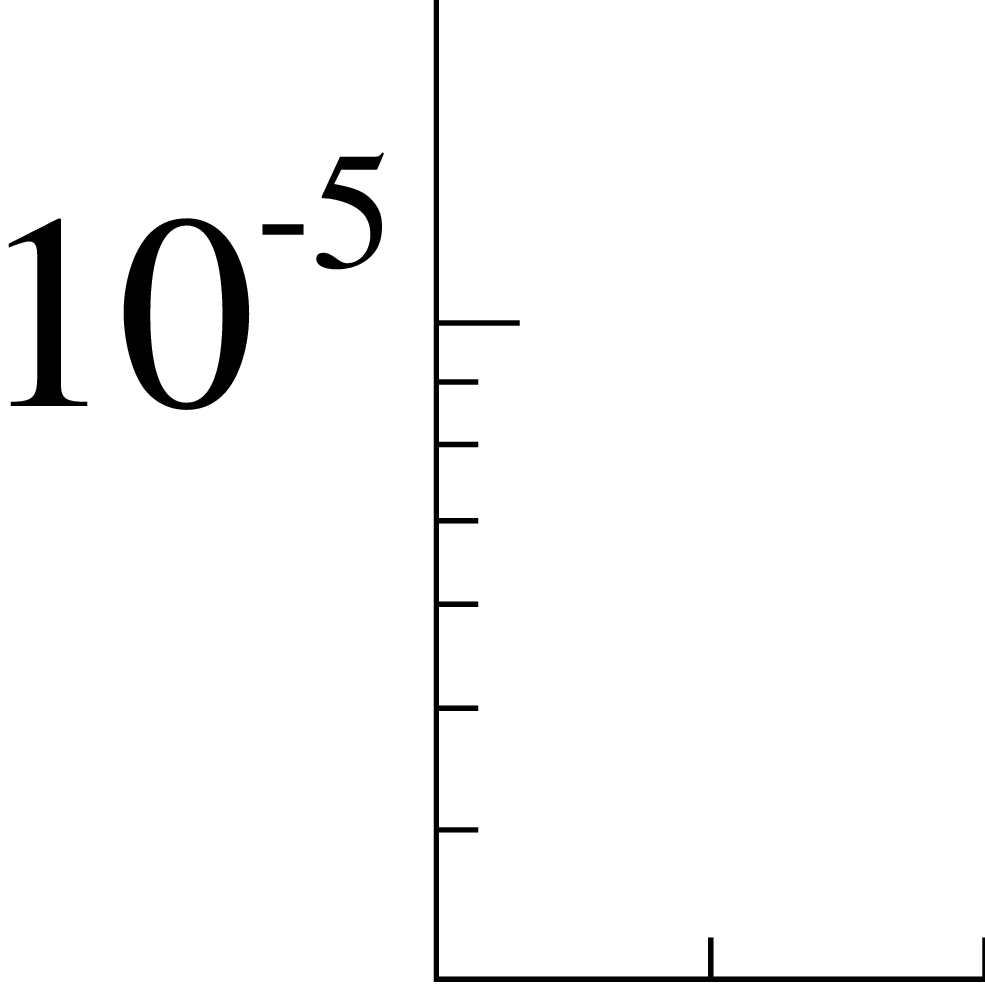}
\end{center}
\caption{Coordination number density for a $65536$ point random lattice.\label{fig.coord}}
\end{figure}

\begin{figure}
\begin{center}
\pic{0.7}{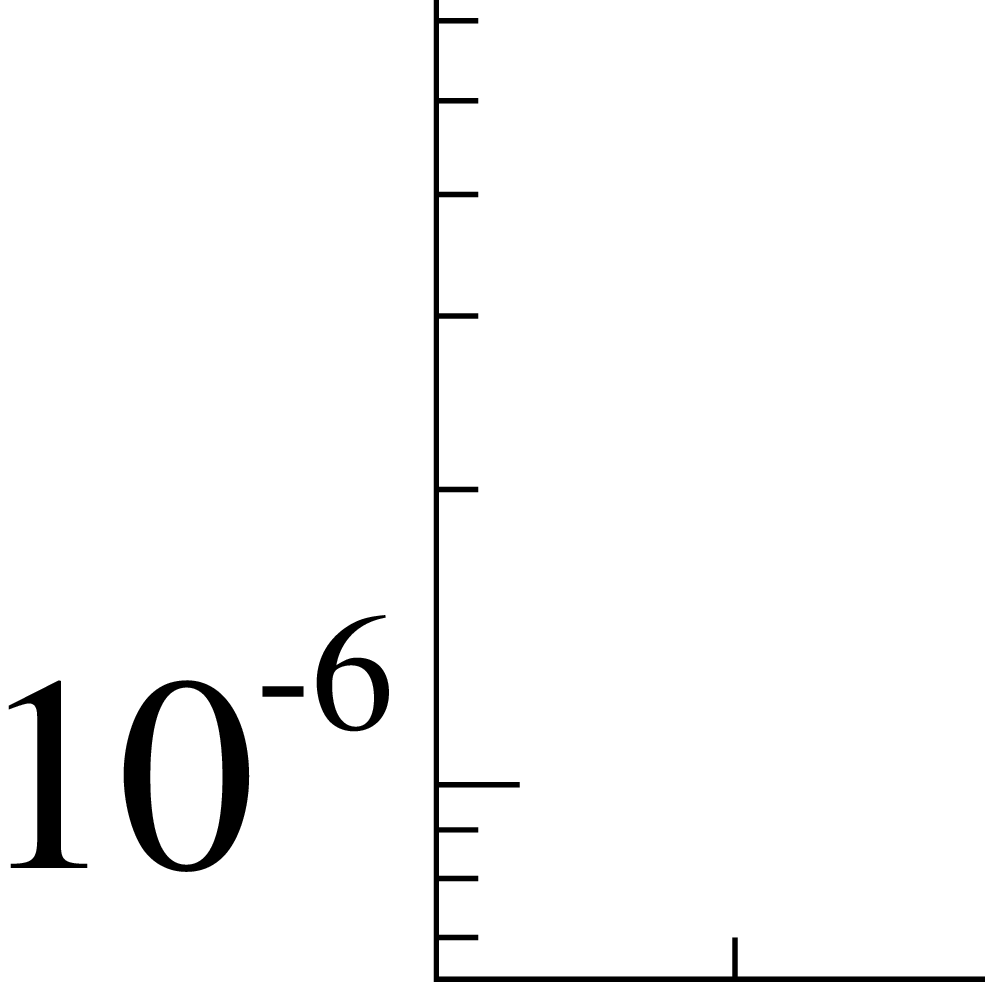}
\end{center}
\caption{\label{fig.lleng}Link length distribution for a $65536$ point lattice.}
\end{figure}

Figure~\ref{fig.coord} details $\rho_N(C)$. Initially $\rho_N(6)=1$ is the only 
non--zero value, but after only one scan, the distribution is approximately 
normal (a Gaussian approximation is shown by the shaded line in figure \ref{fig.coord}), 
and is also very similar to the standard lattice studied as detailed 
in ref.~\cite{drouffe}. The difference is notably that our lattice
exhibits a broader distribution of coordination numbers, with significantly 
more vertices of high coordination number.
Further scans produce more broadening, with even more
vertices of high coordination number formed, until 
an equilibrium situation is established where asymptotically,
\begin{equation}
\rho_N(C)\propto e^{-\alpha C}.
\end{equation}
Too many highly coordinated vertices is undesirable particularly
on a fixed vertex lattice, since this encourages long links which reduce the degree of locality of the lattice. With this increased graininess 
one would expect a greater uncertainty in calculations. 
However, this uncertainty is expected to vanish in the large lattice limit since the coordination number distribution has the same form after an equivalent number of scans. ie: it is dictated by local lattice properties.
The link length distribution of the standard lattice can easily be computed from
the formalism presented in ref.~\cite{ran},
\begin{equation}
\rho_L(x)=2\pi^2 a^{-4} x^3 e^{-\pi x^2 a^{-2}}
\end{equation}
our lattice has very different characteristics, as shown in figure \ref{fig.lleng}. The link distribution is approximated by
\begin{equation}
\rho_L(x)=s^{-1}e^{-xs^{-1}}.
\end{equation}
with many more long links than the CFL lattice, as expected. If there are problems with link variable specification, this lattice will emphasise these problems.

\begin{figure}
\begin{center}
\pic{0.6}{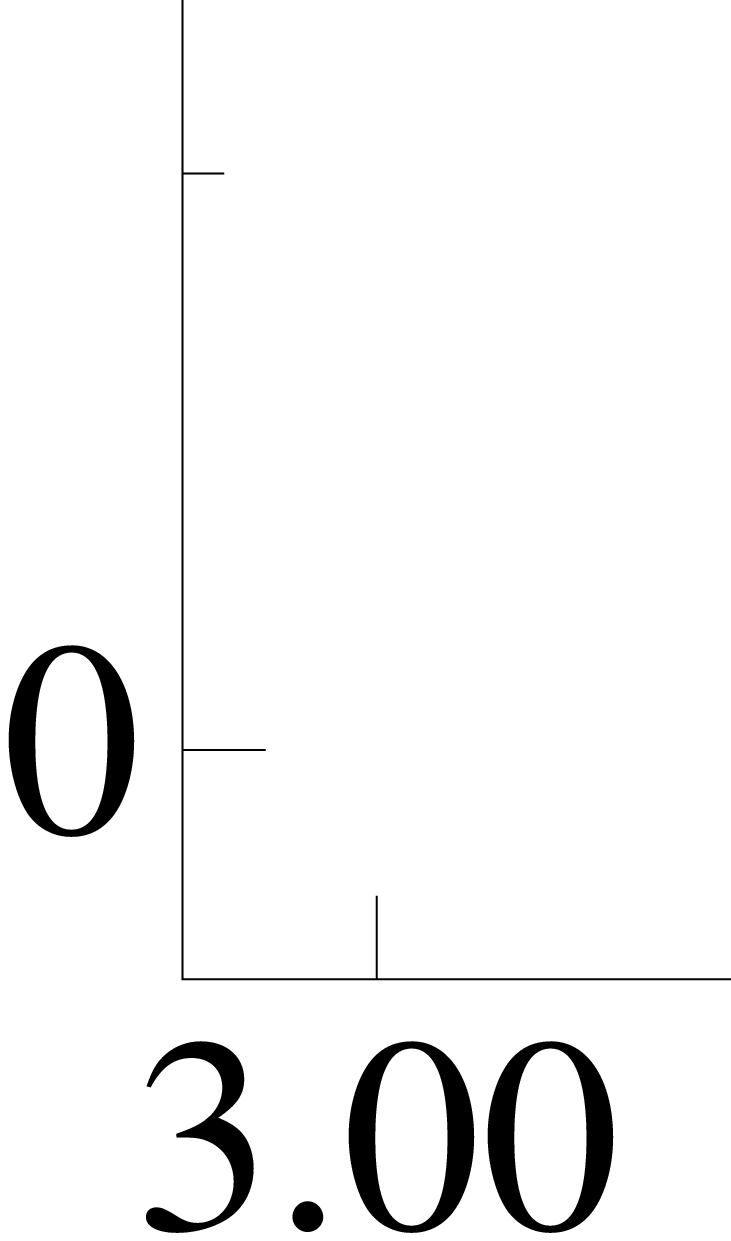}\\
\ref{fig.spin}a. Magnetisation\\
\vspace{5mm}
\pic{0.6}{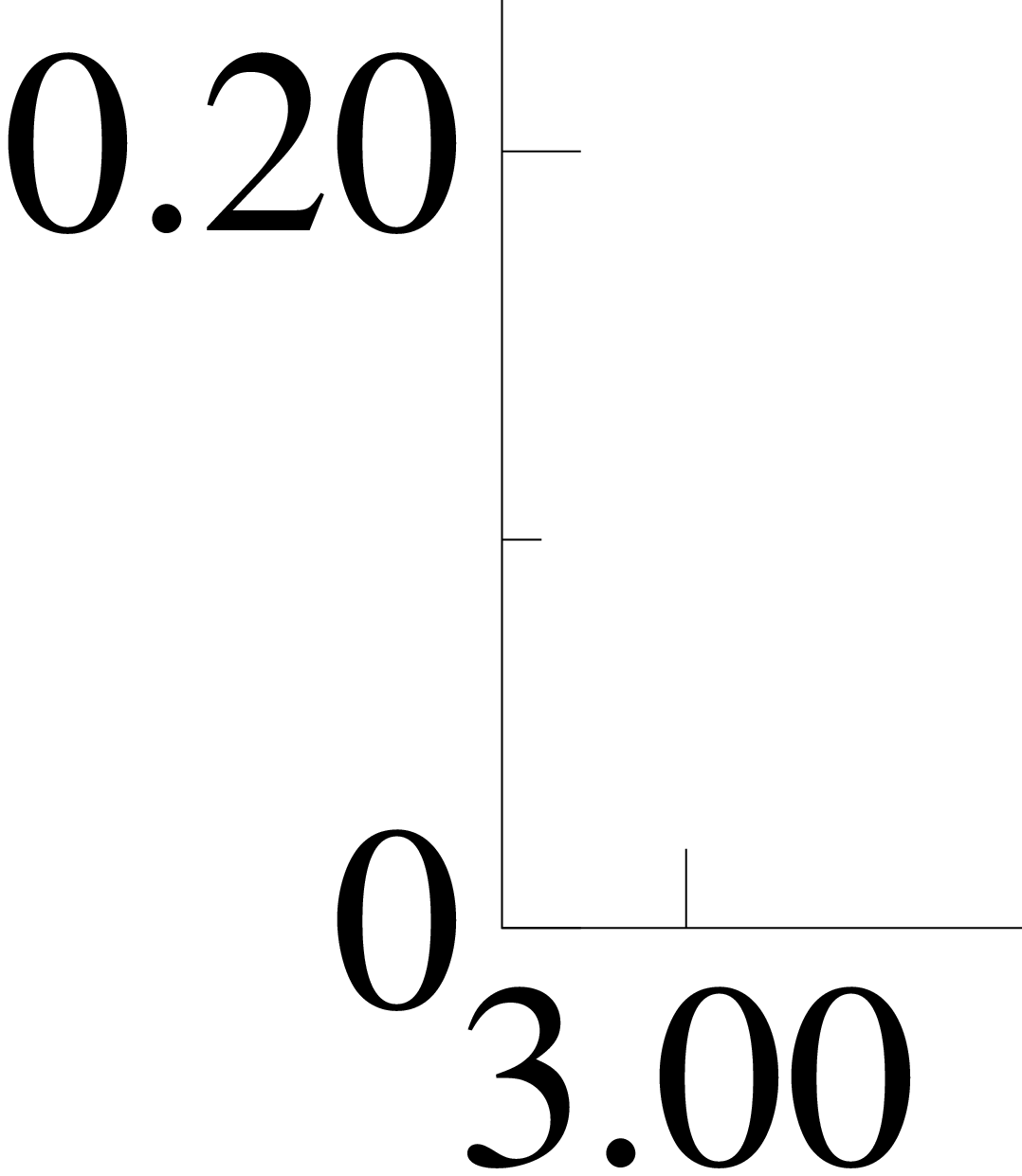}\\
\ref{fig.spin}b. Susceptibility
\end{center}
\caption{Ising spins on our random lattice\label{fig.spin}}
\end{figure}

To further the comparison, we put Ising spins on the vertices of the lattice, with each bond having equal strength couplings. Since the coordination 
number distributions are similar, we do not expect too dissimilar critical behaviour.
The magnetic susceptibility and magnetisation are calculated from 5000 spin configurations of a $65536$ vertex lattice using the Swendsen--Wang clustering method \cite{swendsen}, 
over a range of $\beta$ either side of criticality. Passing over the 
critical point from either direction indicates no detectable hysteresis,
see figure \ref{fig.spin}. The critical point for one scan is at $\beta=0.2594\pm0.002$, determined by averaging the values determined by
the linear fits in this figure. This compares with $\beta=0.2631$ for the CFL lattice \cite{ising} and
$\beta=\frac{1}{4}\log3=0.2746$ for an unflipped lattice.
As previously alluded to, our lattice does not seem to produce as clean a signal as CFL which has only 1000 points compared to 65536 of ours, the scaling region in particular is
smaller, and the flipping process worsens the behaviour considerably;
$\beta_c$ is reduced, and the critical
exponents are more susceptible to finite size effects, which results in a smaller
scaling region. 
This should not be such a problem for the critical phenomena 
of the fermion theory which naturally suppresses long links (thus, highly connected vertices) with massive propagator.
Another obviously different behaviour is the differing gradients on 
each side of the critical point. 
The results show that the lattices are quite probably in the same universality class as the square lattice, and certainly not in the same class as
the fixed vertex construction of ref.~\cite{dotsenko}.

Having briefly investigated the lattice construction procedure, we now
consider the fermion action on this lattice.
\section{The Fermion Action}

The (Euclidean) fermion action is derived from the continuum in the
most direct way
\begin{eqnarray}
S&=&\sum_{x}\;\frac{1}{2}\left(\sum_{l}\;\overline\Psi_{x}\;\gamma_\mu\;\Delta_\mu^{x,+{l}}\;\Omega_{x,x+l}\;\Psi_{x+{l}}\right)
-\sum_x\;\frac{1}{2}\left(\sum_{l}\;\overline\Psi_{x+{l}}\;\gamma_\mu\;\Delta_\mu^{x,-{l}}\;\Omega_{x+l,x}\;\Psi_{x}\right)\nonumber\\
&&+\;m\;\sum_x\Omega_{x,x}\;\overline\Psi_x\;\Psi_x,
\end{eqnarray}
where $\Delta_\mu$ is the lattice derivative, and $\Omega_{x,y}$ weights the volume contribution of the vertices at each end of a link.
Both are chosen in a manner which reduces to the naive result on regular lattices
 \cite{wilson,reglat}.
At a vertex $x$ with coordination number $C_x$, the derivative
is constructed by averaging the contributions
of pairs of orientation-consecutive links $\{({k},{l})\}$.

For a pair of such non-parallel links,
\begin{equation}
\left(\begin{array}{c}\partial_0\Psi(x)\\\partial_1\Psi(x)\end{array}\right)\approx
\left(\begin{array}{cc}k_0&k_1\\l_0&l_1\end{array}\right)^{-1}
\left(\begin{array}{c}\Psi(x+k)-\Psi(x)\\
\Psi(x+l)-\Psi(x)\end{array}\right)
\end{equation}
The lattice construction ensures the existence of the inverse
for all pairs of links which belong to the same simplex. Since each 
simplex should be treated on an equal footing, we form the 
average contribution over all such pairs which surround the vertex,
\begin{eqnarray}
\sum_{l}\;\overline\Psi_{x}\;\gamma_\mu\;\Delta_\mu^{x,+{l}}\;
\Psi_{x+{l}}\;\Omega_{x,x+l}&=&\nonumber\\
\quad\quad\overline\Psi_{x}\;\sqrt{\omega_x}{C^{-1}_x}\sum_{\{({k},{l})\}}
\frac{\gamma}{{k}\times {l}}
&\times&\left[{l}\,\Psi_{x+{k}}\sqrt{\omega_{x+k}}-
{k}\,\Psi_{x+{l}}\sqrt{\omega_{x+l}}+({k}-{l})\Psi_x\sqrt{\omega_x}\right],
\label{eq.deriv}
\end{eqnarray}
$\omega_x=\Omega_{x,x}$ is determined by taking $1/3$ of the area of all triangles which have $x$ as a vertex, the simplest way of discretising the volume
integral. The diagonal $\overline\Psi_x\Psi_x$ term in equation (\ref{eq.deriv}) is canceled by an identical contribution which
emerges when the action is Hermitiansed.

Gauge interactions are introduced in the usual gauge-invariant manner using the link variables
$U_{x,x+{l}}=\exp(ig\int_{l}A(x)\,.\,{\rm d}x)$.
An alternative formulation $U_{x,x+{l}}=\exp(ig\,{l}\,.\,A(x+{l}/2))$, which
is not gauge covariant under the usual continuous gauge transformations, 
is also considered. Note that the long link property of our lattice is
fortuitous in emphasing the difference between these two formulations. The resulting action is hermitian in the Euclidean sense, local, and
apart from the mass term, axially-symmetric.

\section{Suppression of doublers in the free-field case.}

Following ref.\ \cite{espiru},
we first compute a quantity derived from the free propagator
\begin{figure}\begin{center}
\pic{0.7}{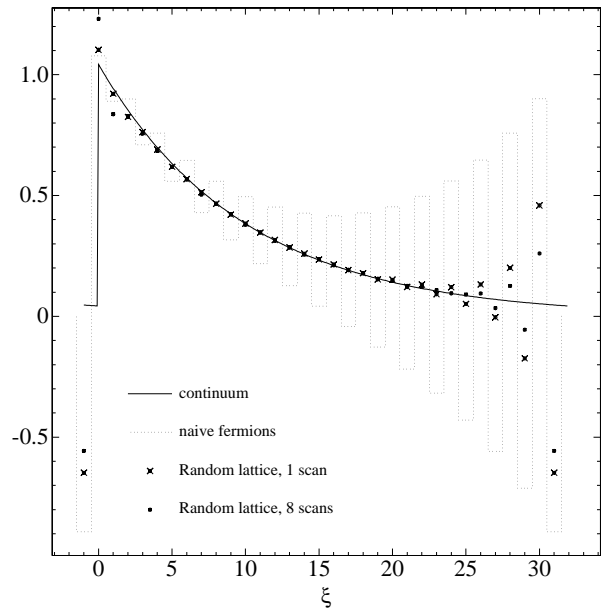}
\end{center}
\caption{\label{fig.fpg} Fermion propagation, $f(\xi)$.}
\end{figure}
\begin{equation}
f(\xi)\ ={\rm Tr_\gamma}\,\frac{1}{a^2N}\int_{x,x'}(1+\gamma_0)\,G_0^{-1}(x,x')\,\delta^1(x_0-x'_0-\xi),
\end{equation}
evaluating the average zero momentum real particle propagator projected along the $x_0$ direction.
In the continuum theory with toroidal boundary conditions, this can
be simply computed from the free continuum fermion propagator on a regular torus
of area $V$,
\begin{equation}
G_0^{-1}(x)=\frac{1}{V}\sum_{k=\{\frac{q\pi}{V^{1\over2}},\frac{p\pi}{V^{1\over2}}\}}\frac{\gamma_\mu k_\mu+m}{{k}^2+m^2}e^{i{k}\cdot{x}},
\end{equation}

\begin{equation}
f(\xi)=V\frac{e^{-m\xi}}{1-e^{-mV^{1\over2}}}
\end{equation}
Fig.~\ref{fig.fpg} indicates the results with naive fermions on a square lattice, and two random lattices, as well as the continuum, clearly identifying the doubler
suppression of free fermions on the random lattice in agreement with \cite{espiru}.
Indeed, apart from some small
distance $O(s)$ deviations in the real particle propagation, 
the doublers are well suppressed at large distances where the 
random lattice result matches the continuum
completely in both normalisation and mass.
The small distance fluctuation of the real mode is particularly apparent on the eight scan lattice, which also shows a higher degree of doubler suppression. Thus,
our lattice agrees with the conventional picture of the suppression of fermion 
doubling in the free-field case on a random lattice. 

\section{Revival of doubling}
We consider the specific background Abelian gauge field,
\begin{eqnarray}
g A_\mu&=&\delta_{\mu,1}\frac{E\sqrt{N}}{2\pi a}\cos\left(\frac{2\pi x_0}{a \sqrt{N}}\right),
\end{eqnarray}
with fixed physical quantities:
${\rm mass}=m=0.1$, ${\rm area}=N=64$, and ${\mbox{\rm electric field}}=E=0.05$, for 
${\mbox{\rm lattice spacing}}=a=\{1.0,0.5,0.3333,0.25\}$, in both 
covariant and non--covariant link variables.

The different choice of link variables amounts to $g\rightarrow g\sin(l_0\pi N^{-1/2})N^{1/2}/(l_0\pi)$, a perturbative effect of $\sim2$\% on a $64$ vertex ($a=1$) lattice and $\sim0.1$\% on $1024$ vertices ($a=0.25$). It should be reiterated that although apparently 
small, this correction is the difference between exact gauge invariance 
and no gauge invariance at finite lattice spacing, which has significant consequences.

The calculation of $\Gamma_2$ on a random lattice is complicated by its sensitivity to
the structure of the lattice.
This sensitivity can be considered as having two sources: variations in $s$,
and the variation due to
inequivalent arrangements of links which have similar $s$.
\begin{figure}\begin{center}
\pic{0.7}{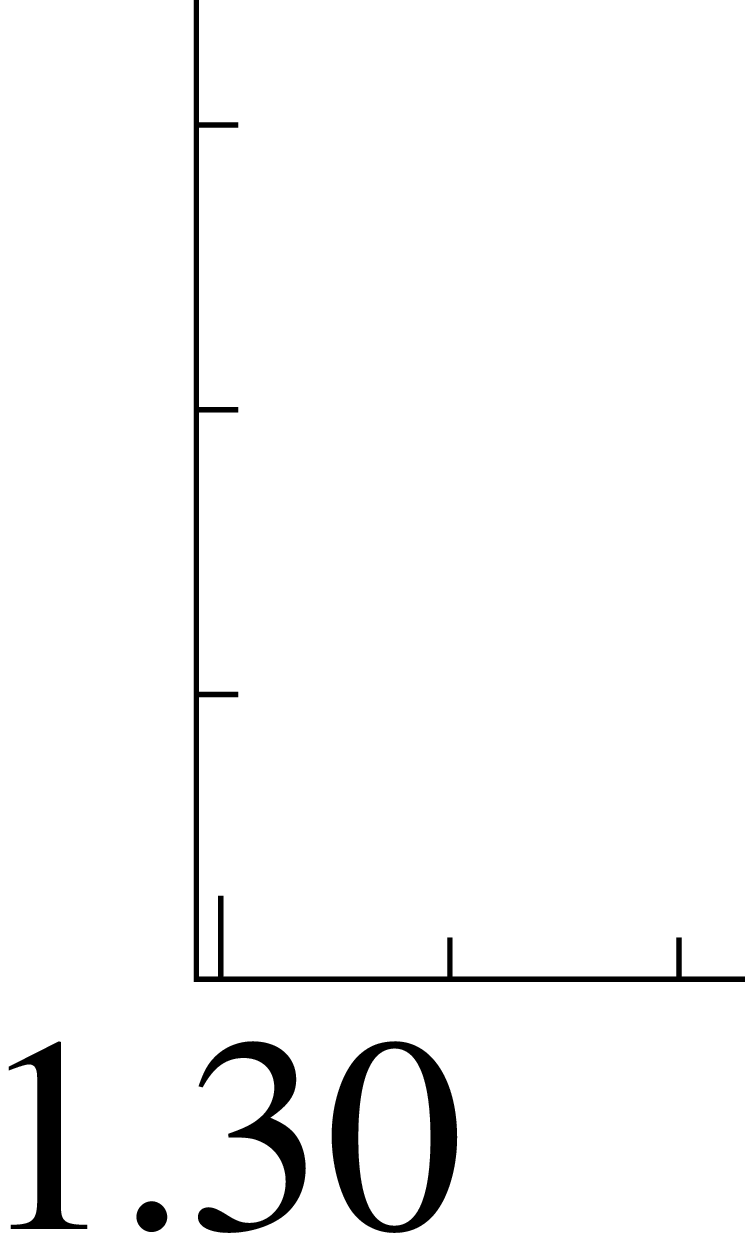}
\end{center}
\caption{\label{fig.measure}The two-point function for fixed physical quantities and lattice size, varying $s$.}
\end{figure}
The later variation gives some estimate of the 
uncertainties associated with non--zero $a$; if it were too large, then it might
be difficult to conclude anything about the number of species present. 
To account for both of these 
we consider an ensemble of lattices randomised by $1\ldots8$ scans. 
Figure~\ref{fig.measure} shows $\Gamma_2$ for both gauge covariant and non-covariant formulations.
The gauge invariant formulation gives a result which is always
at least four times greater than the gauge non--invariant case.
Variation with $s$, and fluctuations around a fixed $s$ are both apparent,
even with such a small sample.
There also appears to be a clustering of the results along two bands in the
gauge invariant case. This is
clear in the $1024$ vertex data presented here in figure \ref{fig.measure}, although not so evident on smaller lattices due to large fluctuations. It could indicate the discrete change in number of species which characterises lattice doubling behaviour. The trends $\Gamma_{2,\rm invariant}$ increasing 
and $\Gamma_{2,\rm non\ invariant}$ decreasing with increasing $s$
are also seen on the smaller lattices. For the purpose of
comparison, we ignore the band structure, and linearly extrapolate to $s=a$, which provides a lower bound on the ratio between the two formulations.
\begin{figure}\begin{center}
\fplotsq{0.7}{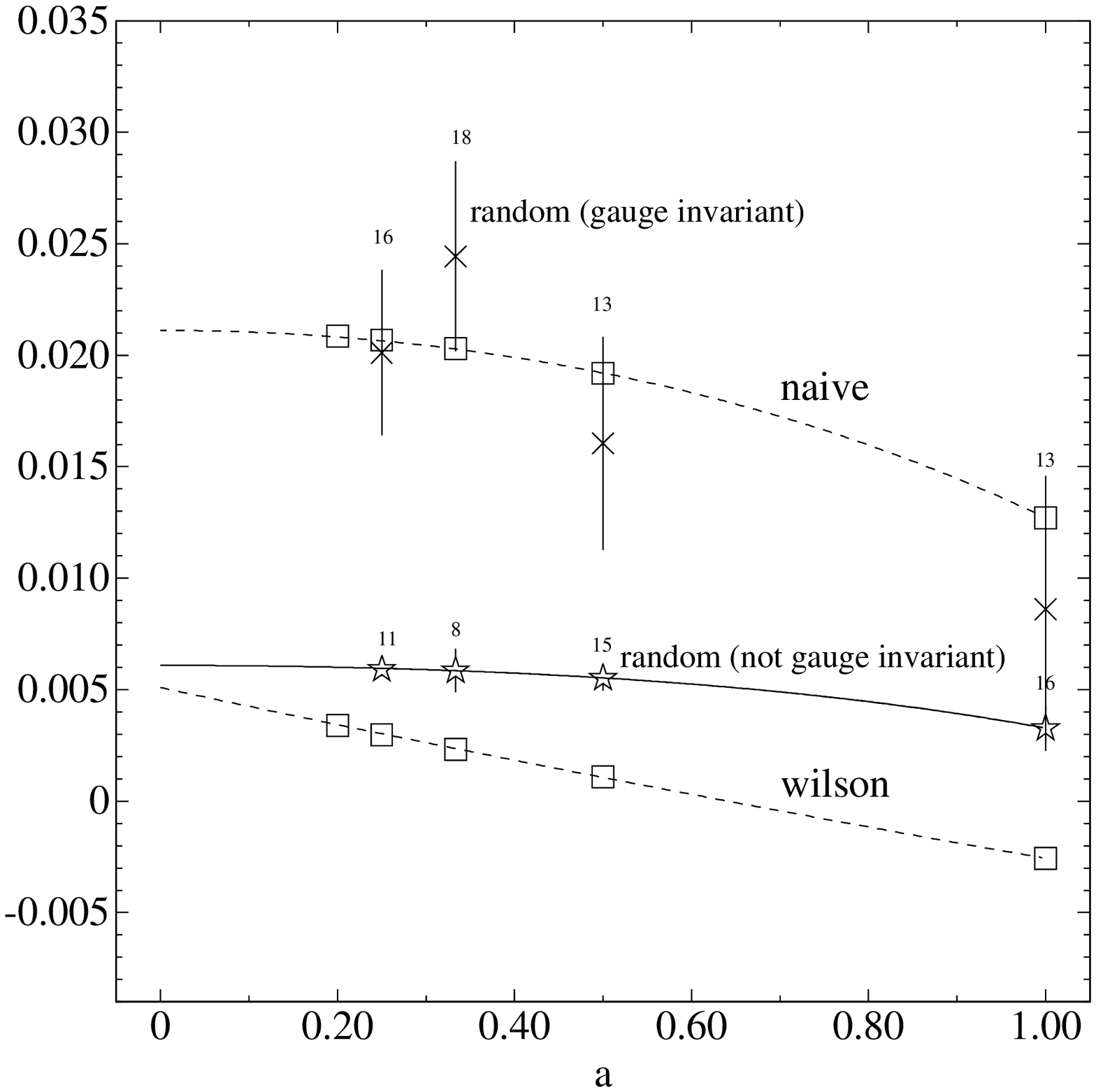}\end{center}
\caption{\label{fig.loop}The two-point function for fixed physical quantities, varying $a$.}
\end{figure}
Identical calculations with naive and Wilson fermions on square lattices of the same sizes
are also included for comparison. See Fig.~\ref{fig.loop}.
The naive case approaches the continuum limit quadratically with $a$ and the Wilson approaches
$1/4$ of the same result linearly, as expected.
The random lattice results include the number of lattice configurations used in the extrapolation next to each point.
The gauge invariant result is clearly more like the naive fermion than the Wilson. 
With gauge invariance broken, the converse is clearly
seen; the result is certainly more like Wilson than naive. Large fluctuations
in the gauge invariant calculations are expected, since if there are spurious modes, the number of modes will be sensitive to the detailed lattice structure,
which we have no control over.
It is also clear that the gauge non-invariant formulation approaches 
the continuum result more rapidly than either Wilson or naive formulations, 
as expected using a random lattice approach.

\section{Discussion}

It is clear from our results that there are doublers on random
lattices when gauge invariance is maintained at finite lattice spacing,
since the extrapolated determinant is comparable to that of naive
fermions. 

  It can also be seen that the doubling can be avoided if one
gives up gauge invariance on the lattice. The hope is that 
it will be recovered again in the continuum limit. This is 
certainly true naively, but must be considered more carefully with chiral gauge
interactions.

In all cases, the lattice fermion actions are invariant under the
global axial transformations.  When there are doublers on random lattices,
the axial anomalies are canceled in the usual manner among opposite-chirality
species.  When there is no doubling in the gauge non-invariant formulation,
the conserved lattice current being the N\"oether current of axial
symmetry is not gauge invariant.  Thus it cannot be
identified with the continuum axial current, and should instead be
identified with a combination of the continuum
current and a gauge-noninvariant term, whose divergence gives us the
axial anomalies,
\begin{eqnarray}
J^{5\mu}_{\rm lattice}(x) &=& J^{5\mu}_{\rm continuum}(x) -
\frac{g}{2\pi}\epsilon^{\mu\nu}A_\nu(x).
\end{eqnarray}

We believe that the results obtained here are also applicable to other
kinds of random lattices in so far as translational invariance is broken. 
The calculations of
\cite{chiu} seem to support this claim, even though the interpretation
and thus conclusions reached there are different to ours.
Previous approaches remove the doublers by point splitting methods \cite{chiu}
or introducing naive vertex operators in the calculation \cite{espiru}, 
which are inconsistent with the Ward--Takahashi identities.
Such aids explicitly break gauge invariance and thus, in the light of our calculation, it is not surprising that doubling may be removed. 
Indeed giving up gauge invariance at some scale on par with the lattice
dimensions 
is also addressed indirectly in remarks about the general problem
of overlocalisation in some physical theories in ref.~\cite{finescale}.
Our calculations are not inconsistent with these ideas.

Our doubling conclusion for random lattices is
not plainly disappointing but also points to some serious implications.

The lattice no-go theorem has thus been extended, and the importance
of gauge invariance emphasised in the phenomenon of
lattice fermion doubling.
The failure of random lattices to accommodate chiral fermions may
undermine the point of view that at the Planck scale or higher the
structure of spacetime is that of randomness;
or may indicate that there is a deep connection between the structure and scale of
space time and gauge invariance; or, taken with other
complete failures in dealing with chiral fermions, could be a hint that
our understanding of chiral gauge theories is incomplete.  
Correspondingly, the quantisation of those theories is in need of further
studies.  One of us has been pursuing this latter path~\cite{tdkcgt}.

\bigskip\noindent{\large\bf Acknowledgments}\medskip

\noindent We acknowledge Dr Lloyd Hollenberg, Dr Girish Joshi and Prof Bruce
McKellar for their interests and support and Prof Tetsuyuki
Yukawa for discussions and guidance on random lattices.  TDK wishes to
thank Dr John Wheater and the Edinburgh Theory Group for discussions,
and acknowledges the support of an Australian Research
Council Fellowship and the
Pamela Todd Award.

\end{document}